\documentclass[10pt,]{article}
\usepackage{lmodern}
\usepackage{amssymb,amsmath}
\usepackage{ifxetex,ifluatex}
\usepackage{fixltx2e} 
\ifnum 0\ifxetex 1\fi\ifluatex 1\fi=0 
  \usepackage[T1]{fontenc}
  \usepackage[utf8]{inputenc}
\else 
  \ifxetex
    \usepackage{mathspec}
    \usepackage{xltxtra,xunicode}
  \else
    \usepackage{fontspec}
  \fi
  \defaultfontfeatures{Mapping=tex-text,Scale=MatchLowercase}
  
\fi
\IfFileExists{upquote.sty}{\usepackage{upquote}}{}
\IfFileExists{microtype.sty}{%
\usepackage{microtype}
\UseMicrotypeSet[protrusion]{basicmath} 
}{}
\usepackage[margin=1in]{geometry}
\usepackage{natbib}
\usepackage{color}
\usepackage{fancyvrb}

\DefineVerbatimEnvironment{Highlighting}{Verbatim}{commandchars=\\\{\}}
\newenvironment{Shaded}{}{}
\newcommand{\KeywordTok}[1]{\textcolor[rgb]{0.00,0.00,1.00}{{#1}}}
\newcommand{\DataTypeTok}[1]{{#1}}
\newcommand{\DecValTok}[1]{{#1}}

\newcommand{\FloatTok}[1]{{#1}}

\newcommand{\StringTok}[1]{\textcolor[rgb]{0.00,0.50,0.50}{{#1}}}

\newcommand{\OtherTok}[1]{\textcolor[rgb]{1.00,0.25,0.00}{{#1}}}

\newcommand{\NormalTok}[1]{{#1}}
\usepackage{graphicx}
\makeatletter
\def\maxwidth{\ifdim\Gin@nat@width>\linewidth\linewidth\else\Gin@nat@width\fi}
\def\maxheight{\ifdim\Gin@nat@height>\textheight\textheight\else\Gin@nat@height\fi}
\makeatother
\setkeys{Gin}{width=\maxwidth,height=\maxheight,keepaspectratio}
\ifxetex
  \usepackage[setpagesize=false, 
              unicode=false, 
              xetex]{hyperref}
\else
  \usepackage[unicode=true]{hyperref}
\fi
\hypersetup{breaklinks=true,
            bookmarks=true,
            pdfauthor={Oleg Sofrygin (sofrygin@berkeley.edu, Department of Biostatistics, University of California, Berkeley, USA); Romain Neugebauer (Division of Research, Kaiser Permanente Northern California, Oakland, USA); Mark J. van der Laan (Department of Biostatistics, University of California, Berkeley, USA)},
            pdftitle={Conducting Simulations in Causal Inference with Networks-Based Structural Equation Models},
            colorlinks=true,
            citecolor=blue,
            urlcolor=blue,
            linkcolor=magenta,
            pdfborder={0 0 0}}
\let\rmarkdownfootnote\footnote%
\def\footnote{\protect\rmarkdownfootnote}

\usepackage{titling}

  \title{Conducting Simulations in Causal Inference with Networks-Based
Structural Equation Models}
  \pretitle{\vspace{\droptitle}\centering\huge}
  \posttitle{\par}
  \author{Oleg Sofrygin
(\href{mailto:sofrygin@berkeley.edu}{\nolinkurl{sofrygin@berkeley.edu}},
Department of Biostatistics, University of California, Berkeley, USA) \\ Romain Neugebauer (Division of Research, Kaiser Permanente Northern
California, Oakland, USA) \\ Mark J. van der Laan (Department of Biostatistics, University of
California, Berkeley, USA)}
  \preauthor{\centering\large\emph}
  \postauthor{\par}
  \predate{\centering\large\emph}
  \postdate{\par}
  \date{2017-05-29}

\usepackage{amsmath}
\usepackage{graphicx}
\usepackage{float}
\usepackage{booktabs}
\usepackage{ctable}
\let\code=\texttt
\let\proglang=\textsf
\newcommand{\pkg}[1]{{\fontseries{b}\selectfont #1}}

\begin{document}

\maketitle

\begin{abstract}
The past decade has seen an increasing body of literature devoted to the
estimation of causal effects in network-dependent data. However, the
validity of many classical statistical methods in such data is often
questioned. There is an emerging need for objective and practical ways
to assess which causal methodologies might be applicable and valid in
network-dependent data. This paper describes a set of tools implemented
in the simcausal R package that allow simulating data based on
user-specified structural equation model for connected units.
Specification and simulation of counterfactual data is implemented for
static, dynamic and stochastic interventions. A new interface aims to
simplify the specification of network-based functional relationships
between connected units. A set of examples illustrates how these
simulations may be applied to evaluation of different statistical
methods for estimation of causal effects in network-dependent data.
\end{abstract}

\section{Introduction}\label{introduction}

The past decade has seen an increasing body of literature devoted to
estimation of causal effects in network-dependent data. Many of these
studies seek to answer questions about the role of social networks on
various aspects of public health. For example, Christakis et al. used
the observational data on subjects connected by a social network to
estimate the causal effects of contagion for obesity, smoking and a
variety of other outcomes
\citep{christakis2007spread, christakis2013social}, finding that many of
these conditions are subject to \emph{social contagion}, e.g., in one of
the studies the authors found that the person's risk of becoming obese
increases with an additional obese friend, even when one controls for
all \emph{measured} confounding factors. However, the statistical
methods employed by these studies have come under scrutiny due to
possibly anti-conservative standard error estimates that did not account
for network-dependence among the observed units \citep{lyons2011spread},
and possibly biased effect estimates that could have resulted from:
model misspecification \citep{lyons2011spread, vanderweele2013social},
network-induced homophily \citep{shalizi2011homophily}, and unmeasured
confounding by environmental factors \citep{shalizi2011homophily}. Given
the potential high impact of such studies on future policy and public
health decisions, it is important to be able to verify the robustness of
statistical methods employed in these and similar studies. Clearly,
there is an emerging need for methods which can test the validity of
various statistical methods for social network data and we argue that
one of the possible approaches involves the conduct of simulation
studies. That is, a practical way to test the validity of a certain
statistical method involves its application against a large set of
plausible data generating scenarios, specifically in relation to the
types of dependent data one might see in social networks. Moreover, a
carefully designed simulation study can test the method's sensitivity to
violations of its key assumptions, highlight its limitations for
specific types of data, and provide an important proof of concept that
complements the results based on statistical theory. Finally,
simulations can be also helpful in identifying errors in the
implementation of complex statistical algorithms.

While estimation of causal effects in social network settings is
challenging, it is also an active area of ongoing research
\citep{halloran2016}. Some of the challenges for inference are due to
the fact that the frequently made assumption of independence among units
is generally invalid. Consider a setting with single time point exposure
where we collect data on the baseline covariates, the exposures and the
outcomes on \(N\) units. When these units are also connected by a
network we might expect that the interactions between any two connected
units can cause the exposure of one unit to have an effect on the
outcome of the other unit - an occurrence often referred to as
interference or spillover \citep{Sobel2006, hudgens2008toward}.
Moreover, it is possible that both, the outcome and the exposure of one
unit, are dependent on the baseline covariates of other connected units.
This paper is concerned with design and implementation of simulation
studies for these types of network-dependent data. We describe a
comprehensive set of tools implemented in the \pkg{simcausal}\footnote{\pkg{simcausal}
  package was developed using the \proglang{R} system for statistical
  computing \citep{r} and it is available from the Comprehensive
  \proglang{R} Archive Network (CRAN) at
  \url{http://CRAN.R-project.org/package=simcausal}.} \proglang{R}
package \citep{R-simcausal, JSSsimcausal}, which allow simulating the
types of dependent data one might collect on a community connected by a
social or geographical network. We then illustrate potential application
of \pkg{simcausal} by describing a simulation study that assesses the
performance of several estimators of causal effects among
network-dependent observations.

In general, \pkg{simcausal} uses the following two-step process for
specifying the data generating distribution for \(N\) connected units.
First, one describes the network of connections between these units
(e.g., social or geographical network) by specifying either a known
network graph or a probabilistic network graph model for \(N\) nodes.
Next, one specifies the distribution of the unit-level covariates (node
attributes) by parameterizing a structural equation model (SEM) for
connected units \citep{vdL2014nets}. This SEM allows the covariates of
one unit to be dependent on the covariates of other connected units via
some known functional form which is controlled explicitly by the user.
For this purpose, we developed a novel \proglang{R} interface which
simplifies the specification of complex network-based functional
relationships between such units. Moreover, the network-based syntax can
be \emph{combined} with the existing syntax for specifying longitudinal
data structures, allowing for simulations of network-based processes
that also evolve in time. We refer to \citet{JSSsimcausal} for general
information about the \pkg{simcausal} package and the description of its
syntax for simulating longitudinal independent and identically
distributed (IID) data.

Network simulation studies have been previously applied to assess the
validity of different estimation approaches in causal inference (for
example, see \citet{Noel2011}, \citet{eckles2014design} and
\citet{Athey2015}). Such simulation studies have also been used as a
guiding tool for comparison of the benefits of different experimental
design strategies in network settings
\citep{Aral2011, walker2014design, basse2015, harling2016leveraging}.
However, to the best of our knowledge, there is no open-source
simulation software for conducing network-based causal inference
research that is similar to the tools and syntax implemented in the
\pkg{simcausal} package. While our package provides the user with a
broad range of possible data-generating distributions, it is
specifically targeted towards causal inference research on the types of
observational data that might be collected when observing members of a
single social network. In other words, \pkg{simcausal} allows
specification of a model based on a causal Directed Acyclic Graph (DAG),
and in conjunction with previously specified network model, this causal
DAG can be used to define ways in which connected units depend on each
other. Moreover, any \proglang{R} package that can simulate a network
graph can be used within \pkg{simcausal}. For example, in this article
we rely on the \pkg{igraph} \proglang{R} package \citep{igraph} for
simulating various network graphs.

In summary, our newly developed software provides a simple and concise
interface for specifying a network data-generating distribution and
incorporating the network structure into various forms of functional
dependence of one unit on other \emph{connected} units. It permits the
simulation of such interconnected data structures, as well as the
generation of the counterfactual data under single or multiple
time-point interventions. The user-defined causal parameters can then be
evaluated from the counterfactual data and serve as gold-standards in
testing the validity of various statistical methods. Finally, the
\pkg{simcausal} package provides the syntax for specifying complex
longitudinal data structures \emph{in combination} with the
network-dependent syntax, allowing one to simulate complex network
processes that also evolve over time.

\subsection{Other related R packages}\label{other-related-r-packages}

As of May 29, 2016, there were 208 \proglang{R} packages on CRAN that
contain the word ``network'' in their title sentence. A large number of
these packages are dedicated towards visual analysis and representation
of networks (e.g., \pkg{ggnetwork} \citep{R-ggnetwork}, \pkg{igraph}
\citep{igraph}, \pkg{d3Network} \citep{R-d3Network}), re-constructing
the network based on biological, neural and other field-specific data
(e.g., \pkg{interventionalDBN} \citep{R-interventionalDBN},
\pkg{LogitNet} \citep{R-LogitNet}, \pkg{RSNNS} \citep{R-RSNNS},
\pkg{dna} \citep{R-dna}), statistical analysis of networks based on
specific network-generating models (\pkg{multiplex} \citep{R-multiplex},
\pkg{lvm4net} \citep{R-lvm4net}). In addition, a large collection of
packages often referred to as a ``\pkg{statnet} suite of packages''
provides various tools for social network analysis, visualization,
simulation and diagnoses based on the statistical methods of
exponential-family random graph models (ERGMs) or other related
parametric model families for networks (e.g., \pkg{statnet}
\citep{statnetJSS}, \pkg{sna} \citep{R-sna}, \pkg{EpiModel}
\citep{R-EpiModel}, \pkg{ergm} \citep{R-ergm}, \pkg{tergm}
\citep{R-tergm}). Other packages for statistical analysis of network or
network-related data include, among others, \pkg{netdiffuseR}
\citep{R-netdiffuseR}, \pkg{nets} \citep{R-nets}, \pkg{ebdbNet}
\citep{R-ebdbNet} and \pkg{tmlenet} \citep{R-tmlenet}. Finally, a large
number of \proglang{R} packages are targeted towards analyses and
simulation of various networks, network evolution over time and the
modeling various network features, such as the creation of the tie
between two nodes and answering questions such as, how and why certain
network ties are formed? Among the packages that are tailored to such
analyses are \pkg{CIDnetworks} \citep{R-CIDnetworks}, \pkg{tsna}
\citep{R-tsna}, \pkg{networkDynamic} \citep{R-networkDynamic} and
\pkg{egonet} \citep{R-egonet}. Another class of packages that is worth
noting are those specifically targeted towards modeling of epidemics on
a network graph, such as packages \pkg{epinet} \citep{R-epinet},
\pkg{netdiffuseR} \citep{R-netdiffuseR}, \pkg{EpiModel}
\citep{R-EpiModel}, for example, by relying on the agent-based modeling
or ERMG techniques. In addition, the following packages are specifically
designed for simulations of the network-graph data: \pkg{SocialMediaLab}
\citep{R-SocialMediaLab}, which provides tools for collecting and
generating social media data for networks; \pkg{hybridModels}
\citep{R-hybridModels}, which allows simulations of stochastic models
for transmission of infectious diseases in longitudinal networks;
\pkg{NetSim} \citep{R-NetSim}, a package for simulating social networks;
and finally, \pkg{RSiena} \citep{R-RSiena}, a package designed for
simulation-based analysis of networks as well as model fitting for
longitudinal network data.

\subsection{Organization of this
article}\label{organization-of-this-article}

The rest of this article is organized as follows. In Section \ref{SEM},
we formally describe our assumed observed data structure and provide the
technical definition of the underlying structural equation model (SEM)
for possibly connected units. In Section \ref{interface} we provide some
technical details of the \pkg{simcausal} interface for simulating data
according to the user-specified parameterization of this SEMs. In
Section \ref{sim.Example1}, we illustrate the use of \pkg{simcausal} in
a typical simulation study, evaluating the true sample-average mean
causal outcome (\emph{the gold-standard}) of a single time-point
stochastic intervention on continuous exposure. We also demonstrate how
this network simulation examples can be used in practice for assessing
the performance of statistical methods for estimation of causal effects
among network-dependent observations. We conclude with a discussion in
Section \ref{discussion}.

\section{Structural Equation Model for Connected
Units\label{SEM}}\label{structural-equation-model-for-connected-units}

We start by introducing some notation. Suppose that we can simulate a
sample of \(N\) connected observations, where each observed unit is
indexed as \(i=1,\ldots,N\). We let
\(F_i \subseteq \{ 1,\ldots,N \} \backslash i\) denote the set of
observations that are assumed ``\emph{connected}'' to \(i\) or, as we
will refer to it, the units in \(F_i\) are ``\emph{friends}'' of \(i\).
In other words, we assume that each set \(F_i\) consists of a unique set
of indices \(j\) in \(\{1,\ldots,N\}\), except for \(i\) itself. We also
allow \(F_i\) to be empty, which would imply that observation \(i\) is
not receiving input from any other units. We assume that \(i\) may
receive input from other observations only if those observations are
listed as part of \(F_i\). We will refer to the union of all \(F_i\) as
a ``\emph{network profile}'' on all \(N\) observations, which will be
denoted by \(\mathbf{F}\). Let \(O_{i}=(W_{i},A_{i},Y_{i})\) denote the
data collected on each observation \(i\), for \(i=1,\ldots,N\), where
\(W_{i}\) denotes all the baseline covariates for \(i\), \(A_{i}\)
denotes the exposure of \(i\) and \(Y_{i}\) denotes the outcome of
\(i\). Let \(\mathbf{W}=(W_{i})_{i=1}^{N}\),
\(\mathbf{A}=(A_{i})_{i=1}^{N}\), \(\mathbf{Y}=(Y_{i})_{i=1}^{N}\) and
\(\mathbf{O}=(\mathbf{W},\mathbf{A},\mathbf{Y})\). Finally, we assume
\(F_{i}\in W_{i}\), that is, the network of friends of \(i\) is assumed
to be part of \(i\)'s baseline covariates. Below we define the
structural equation model (SEM) \citep{Pearl2009} for \(N\) connected
units, similar to that presented in \citet{vdL2014nets}. This model will
form the basis for describing the types of network-based data generating
distributions that are considered in this article.

\subsection{General Structural Equation Model for Connected
Units}\label{general-structural-equation-model-for-connected-units}

Suppose that \(N\) connected observations in \(\mathbf{O}\) are
generated by applying the following structural equation model (SEM):
first generate a collection of exogenous errors
\(\mathbf{U}_{N}=(U_{i}:i=1,\ldots,N)\) across the \(N\) units, where
the exogenous errors for unit \(i\) are given by \[
U_{i}=(U_{W_{i}},U_{A_{i}},U_{Y_{i}}),\ \ i=1,\ldots,N,
\] and then generate \(\mathbf{O}\) deterministically by evaluating
functions as follows:

\begin{eqnarray}
W_{i} & = & f_{W_{i}}(U_{W_{i}}),\mbox{ for }i=1,\ldots,N\label{snm-iobs}\\
A_{i} & = & f_{A_{i}}(W_{i},(W_{j}:j\in F_{i}),U_{A_{i}}),\mbox{ for }i=1,\ldots,N\nonumber \\
Y_{i} & = & f_{Y_{i}}(A_{i},W_{i},(A_{j},W_{j}\::\:j\in F_{i}),U_{Y_{i}}),\mbox{ for }i=1,\ldots,N\mbox \nonumber
\end{eqnarray}

\begin{figure}[H]

{\centering \includegraphics{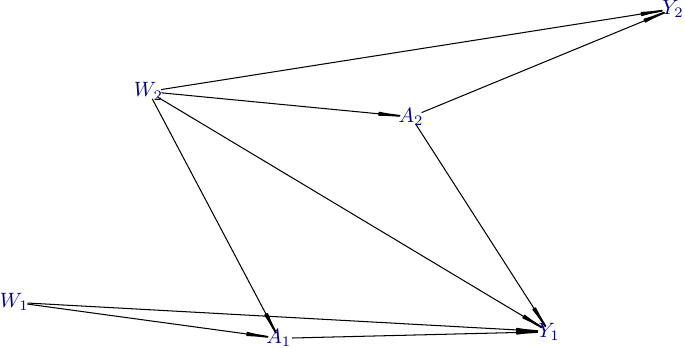}

}

\caption{An example of a directed acyclic graph (DAG) for two observations, where unit 1 is dependent on unit 2, but not vice-versa.}\label{fig:figDAG2depobs}
\end{figure}

The above SEM is general in the sense that it makes no functional
restrictions on \(f_{W_{i}}\), \(f_{A_{i}}\) and \(f_{Y_{i}}\), allowing
each \(i\)-specific set of covariates \((W_i,A_i,Y_i)\) to be generated
from their own \(i\)-specific functions. However, the functions
\(f_{W_{i}}\), \(f_{A_{i}}\) and \(f_{Y_{i}}\) will be subjected to
modeling below. This SEM could be visualized as a causal graph
describing the causal links between the \(N\) units. That is, one could
represent the SEM in \ref{snm-iobs} as a Directed Acyclic Graph (DAG)
\citep{pearl1995}, by drawing arrows from causes to their effects.
However, such a DAG would have to include all \(N\) dependent
observations (the entire network), since the above SEM includes a
separate equation for each observed unit \(i=1,\ldots,N\). Furthermore,
one can imagine that this structural causal model might be also
time-dependent, in which case the causal links among \(N\) units would
evolve over time and the visualization of such structural causal model
would describe what one might call a dynamic causal network
\citep{vdL2014nets}. For example, for a network with two units
(\(N=2\)), in which unit \(i=1\) is dependent on unit \(i=2\), but not
vice-versa, this type of the SEM could be depicted with a DAG shown in
Figure \ref{fig:figDAG2depobs}. We also note that the error terms
\((U_{W_i},U_{A_i},U_{Y_i})\), for \(i=1,2\), are excluded from this
causal DAG \citep{pearl2012causalfoundation}, with the implication that
each of the remaining variables is subject to the influence of its own
error.

\subsection{Dimension Reduction
Assumptions}\label{dimension-reduction-assumptions}

To simplify the notation and to control the dimensionality of the above
SEM, we assume that there are also some known summary measures:
\[W_{i}^{s}:=w_{i}^{s}(W_{i},W_{j}:j\in F_{i})\] and
\[A_{i}^{s}:=a_{i}^{s}((A_{i},W_{i}),(A_{j},W_{j}\::\:j\in F_{i})),\]
and we use the short-hand notation \(W_{i}^{s}\) for
\(w_{i}^{s}(\cdot)\) and \(A_{i}^{s}\) for \(a_{i}^{s}(\cdot)\). We
assume that these summary measures \(W_{i}^{s}\) and \(A_{i}^{s}\) are
of constant-in-\(i\) dimension (that does not depend on \(N\)). Finally,
we assume that for each node \(A_{i}\) and \(Y_{i}\), we can define the
corresponding common (in \(i\)) functions \(f_{A}\) and \(f_{Y}\), so
that

\begin{eqnarray}
W_{i} & = & f_{W_{i}}(U_{W_{i}}),\mbox{ for }i=1,\ldots,N\label{snm}\\
A_{i} & = & f_{A}(W_{i}^{s},U_{A_{i}}),\mbox{ for }i=1,\ldots,N\nonumber \\
Y_{i} & = & f_{Y}(A_{i}^{s},W_{i}^{s},U_{Y_{i}}),\mbox{ for }i=1,\ldots,N.\nonumber
\end{eqnarray}

Examples of such dimension reductions are provided in Section
\ref{interface} and \ref{sim.Example1}. In particular, one could define
\(a_{i}^{s}(\mathbf{W})=(A_{i},(A_{j}\::\:j\in F_{i}))\), i.e., the
observed exposure of unit \(i\) itself and the observed exposures of its
friends. Similarly, one could define
\(w_{i}^{s}(\mathbf{W})=(W_{i},(W_{j}\::\:j\in F_{i}))\). By augmenting
these reductions to data on maximally \(K\) friends, filling up the
empty cells for units with fewer than \(K\) friends with a missing
value, these dimension reductions have a fixed dimension, and include
the information on the number of friends.

\subsection{Independence Assumptions on Exogenous
Errors}\label{independence-assumptions-on-exogenous-errors}

We now make the following (conditional) independence assumptions on the
exogenous errors. First, we assume that each \(U_{W_{i}}\) is
independent of all \(U_{W_{j}}\) such that
\(F_{i}\cap F_{j}=\emptyset\), for \(i=1,\ldots,N\) (weak dependence of
\(\mathbf{W}\)). For example, one of the possible ways to simulate such
dependent \(W_{1},\ldots,W_{N}\) in \pkg{simcausal} is to define a
\emph{latent} (hidden) covariate \(W^*_i\) which is shared between all
observations connected to \(i\) and then defining the \emph{observed}
baseline covariate for \(i\) and all \(j \in F_i\) conditionally on
\(W^*_i\). Such simulation procedure allows one to introduce dependence
between the observed \(W_i\) and \(W_j\), whenever \(i\) and \(j\) are
connected. One could make further restrictions on \(\mathbf{W}\), for
example, by assuming that all \(U_{W_{i}}\) are \emph{unconditionally}
independent, which would imply the independence of \(\mathbf{W}\).
Second, we assume that conditional on \(\mathbf{W}\), all
\((U_{A_{i}}\::\:i=1,\ldots,N)\) are independent and identically
distributed (IID). Finally, we assume that conditional on
\(\mathbf{A}\), all \((U_{Y_{i}}\::\:i=1,\ldots,N)\) are also IID.

The important implication of the latter assumptions is that, given the
observed past on \(N\) baseline covariates and treatments
\((\mathbf{W},\mathbf{A})\), for any two units \(i\) and \(j\) that have
the same value for their summaries \(W_{i}^s=W_{j}^s\) and
\(A_{i}^s=A_{j}^s\), we have that \(Y_{i}\) and \(Y_{j}\) are
independent and identically distributed. Similarly, the same statement
holds for two treatment nodes \(A_{i}\) and \(A_{j}\), whenever the
baseline summaries of \(i\) and \(j\) are equal, i.e.,
\(W_{i}^s=W_{j}^s\). These assumptions on errors
\(U_{i}:=(U_{W_{i}},U_{A_{i}},U_{Y_{i}})\), for \(i=1,\ldots,N\) imply
that: 1. \(W_1,\ldots,W_N\) are (at most) weakly dependent; 2.
Conditional on \(\mathbf{W}\), the random variables \((A_1,\ldots,A_N)\)
are mutually independent; and 3. Conditional on
\((\mathbf{A},\mathbf{W})\), the random variables \((Y_1,\ldots,Y_N)\)
are mutually independent. Thus, all the dependence between units is
explained by the observed pasts of the units themselves and of their
friends. Furthermore, the above structural assumptions will imply
restrictions on the conditional distribution of \(Y_i\) and \(A_i\).
However, the full description of the resulting statistical model based
on the above SEM is outside the scope of this article and we refer the
interested reader to \citet{sofryginJCI2016} for additional details.

The SEM in \ref{snm} and the above independence assumptions define the
set of possible data-generating distributions that are considered in
this article. In particular, one could utilize the \pkg{simcausal}
package to simulate \(N\) dependent observations \(\mathbf{O}\) based on
the following algorithm:

\begin{enumerate}
\def\labelenumi{\arabic{enumi}.}
\itemsep1pt\parskip0pt\parsep0pt
\item
  Start by generating \(N\) baseline covariates \((W_1,\ldots,W_N)\), by
  first drawing \(N\) weakly dependent errors
  \((U_{W_{i}}:i=1,\ldots,N)\) and then applying the equations
  \(f_{W_{i}}(U_{W_{i}})\), for \(i=1,\ldots,N\);
\item
  Generate \(N\) baseline summaries \((W_1^s,\ldots,W_N^s)\), by
  applying the baseline summary measures \(w_i^s(\cdot)\) to \(W_i\) and
  \((W_j:j \in F_i)\), for \(i=1,\ldots,N\);
\item
  Generate \(N\) exposures \((A_1,\ldots,A_N)\), by first drawing \(N\)
  IID errors \((U_{A_{i}}:i=1,\ldots,N)\) and then applying the common
  equation \(f_{A}(\cdot)\) to each \(W_i^s\) and \(U_{A_{i}}\), for
  \(i=1,\ldots,N\);
\item
  Generate \(N\) exposure summaries \((A_1^s,\ldots,A_N^s)\), by
  applying the exposure summary measures \(a_i^s(\cdot)\) to
  \((A_i,W_i)\) and \((A_j,W_j:j \in F_i)\), for \(i=1,\ldots,N\); and
\item
  Generate \(N\) outcomes \(Y_1,\ldots,Y_N\), by drawing \(N\) IID
  errors \((U_{Y_{i}}:i=1,\ldots,N)\) and then applying the common
  equation \(f_{Y}(\cdot)\) to \((A_i^s,W_i^s)\) and \(U_{Y_{i}}\), for
  \(i=1,\ldots,N\).
\end{enumerate}

\subsection{Counterfactuals, Interventions and the Target Parameter as
the Average Causal Effect
(ACE)}\label{counterfactuals-interventions-and-the-target-parameter-as-the-average-causal-effect-ace}

The SEM in \ref{snm} also implicitly encodes the definition of
counterfactual variables, i.e., variables which would result from some
particular interventions on a set of endogenous variables. For example,
for a particular vector of treatment assignments
\(\mathbf{a}=(a_1,\ldots,a_N)\), we could modify the SEM as follows: \[
\begin{array}{lclc}
W_{i} & = & f_{W_{i}}(U_{W_{i}}), & i=1,\ldots,N,\\
A_{i} & = & a_{i}, & i=1,\ldots,N,\\
Y_{i,\mathbf{a}} & = & f_{Y}(a_{i}^{s}(\mathbf{a},\mathbf{W}),W_{i}^{s},U_{Y_{i}}), & i=1,\ldots,N,
\end{array}
\]

where the equations for \(W_i\) were kept unchanged, each \(A_i\) was
set to \(a_{i}\), and each \(Y_{i,\mathbf{a}}\) denotes the
counterfactual outcome of unit \(i\), under the network intervention on
all \(N\) units that sets \(\mathbf{A}=\mathbf{a}\). In this article, we
will refer to \((\mathbf{W},\mathbf{a},\mathbf{Y}_{\mathbf{a}})\) as
\emph{counterfactual data} for \(N\) units and we define our target
causal parameter as a function of such counterfactual data distribution.
For example, the average treatment effect (ATE) can be simply expressed
as \(E\left[\mathbf{Y}_{\mathbf{1}}-\mathbf{Y}_{\mathbf{0}}\right]\).

\section{Syntax for Simulating Networks and Network-Dependent
Data}\label{syntax-for-simulating-networks-and-network-dependent-data}

\label{interface}

To define the distribution of the data (and to simulate such data),
\pkg{simcausal} uses a \code{DAG} object, which will typically consist
of a collection of individual \emph{nodes} (random variables). Such
nodes are defined each time the user calls the \texttt{node()} function.
That is, each call to \texttt{node()} defines the conditional
distribution(s) of either a single or time-varying node. Collectively,
these \texttt{node()} calls define a single \texttt{DAG} object, which
parametrizes the distribution of the data that the user wants to
simulate. By default, the distribution of any new node for a single
observation \(i\) can be dependent only on \(i\)'s values of the
previously defined nodes, and not the node values of other observations.
As a result, any two observations sampled from such \texttt{DAG} object
will be independent. However, there are many real-life examples when one
wishes to simulate data based on some either a priori known or
hypothesized information about the network of connections between
different observations. The new functionality we describe here allows
the user to specify the distribution of a network graph and then use
that network to define the distribution of covariates for each
observation \(i\) conditionally on the covariates of other observations.
For example, if we assume that the observation \(j\) is part of \(i\)'s
social or geographical network (or as we will call it, the observation
\(j\) is a \emph{friend} of \(i\)), then the nodes of \(j\) are allowed
to influence the distribution of the nodes of \(i\). Moreover, we assume
that the functional form of such dependence can be described by some
user-specified network-based summary measures. For that purpose, we used
the \proglang{R} list subsetting operator ``\texttt{{[}{[}...{]}{]}}''
and re-defined it specifically for indexing and building of such network
summaries based on observations that are friends of \(i\) (e.g.,
``\texttt{Var{[}{[}indx{]}{]}}''). We also note that \pkg{simcausal}
does not allow simultaneous friend references of the same node, that is,
each newly added node can be defined only as a function of the nodes
that have been previously added to the \texttt{DAG} object.

\subsection{Defining the network}\label{defining-the-network}

In order to perform network-based simulations with the \pkg{simcausal}
package, the user has to declare a function which will return a specific
network matrix, and we will refer to any such function as the
\emph{network generator}. In particular, a network generator is any
user-specified function that returns a network profile \(\mathbf{F}\) on
\(N\) observations, defined as a matrix of \(N\) rows, each row \(i\)
containing the set of friends \(F_i\) of observation \(i\), i.e., row
\(i\) consists of a vector of observations from \(\{1,\ldots,N\}\) that
are assumed connected to \(i\). The network generator function should
accept at least one named argument, \code{n}, and it must return a
network matrix with \code{n} rows and \code{Kmax} columns, where
\code{Kmax} stands for a maximal possible number of friends for any
observation. When observation \(i\) happens to have fewer than
\code{Kmax} friends, i.e., \(|F_i|<\)\code{Kmax}, the remainder of the
matrix row \(i\) must by filled with \texttt{NA} (missing) values. Note
that an observation \(i\) is allowed to have no friends, which is
denoted by an empty friend set \(F_i\), in which case the row \(i\) of
the network matrix should only consist of \texttt{NA} (missing) values.

Once such a network generator has been defined, the next step is to
\emph{add} this network to a specific \texttt{DAG} object. This is
accomplished by simply calling the \texttt{network} function, specifying
the name of the network generator as its argument ``\texttt{netfun}''
and adding this network function call to the current \texttt{DAG} object
with the \proglang{R} operator ``\texttt{+}''. In other words, the
\texttt{network} function call defines the network and is added to an
existing \texttt{DAG} object with a syntax ``\texttt{+network(...)}''.
Note that this is identical to the \pkg{simcausal} syntax for adding new
\texttt{node} function calls to a growing \texttt{DAG} object when
defining data for IID observations. This network can then serve as a
backbone for defining the dependent-data structural equation models
within such a \texttt{DAG} object. More specifically, the previously
defined node values of \(i\)'s friends can be now referenced as part of
the new \texttt{node} function calls, defining the conditional
distribution of the node for each observation \(i=1,\ldots,N\). The
examples following this section illustrate this functionality for
various network models.

Finally, we note that the \texttt{network} function can accept any
number of user-specified optional arguments, where each of such optional
arguments must be an evaluable \proglang{R} expression. These
expressions will not be evaluated until the data simulation step, at
which point all of them are passed as arguments to the function that
defines the network generator. Thus, just like the regular \texttt{node}
function expressions, the \texttt{network} function expressions can
refer to any standard or user-specified \proglang{R} functions and can
reference any of the previously defined \texttt{DAG} nodes (i.e.,
\texttt{DAG} nodes that were added prior to \texttt{network} function
call). This feature can be useful when, for example, one wishes to
simulate the network in which the probability of forming a tie between
two units depends on the previously simulated unit-specific variable
values (such as the baseline risk factors on each unit).

\subsection{Using the syntax \texttt{{[}{[}...{]}{]}} for network-based
variable
subsetting}\label{using-the-syntax-...-for-network-based-variable-subsetting}

Following the \texttt{network} function call, subsequent calls to
\texttt{node} function can employ our re-purposed list subsetting
operator ``\texttt{{[}{[}...{]}{]}}'' for indexing the node values of
friends. First, the variable which is to be used for network subsetting
is specified in front of the subsetting operator, e.g.,
``\texttt{A{[}{[}...{]}{]}}''. Second, the friend values of the variable
\texttt{A} are specified by the subsetting index, e.g.,
``\texttt{A{[}{[}1:5{]}{]}}''. This expression will look up the values
of node \texttt{A} for friends indexed from 1 to 5 and it will be
evaluated for all observations \(i=1,\ldots,N\). The specific ordering
of friends is determined by the column ordering of the network matrix
returned by the network generator. Such network-indexing expressions can
be also used as inputs of different \proglang{R} functions, enabling
evaluation of various network-based summaries. For example, the
expression ``\texttt{sum(A{[}{[}1:Kmax{]}{]})}'' will specify a vector
of length \(N\) that will consist of a sum of \texttt{A} values among
all friends of \(i\), for each observation \(i=1,\ldots,N\). This syntax
is fully generalizable towards any function which can operate on
matrices, such as the matrix result of the expression
``\texttt{A{[}{[}1:Kmax{]}{]}}''. Moreover, two of the commonly used
\proglang{R} functions, \texttt{sum} and \texttt{mean}, are
automatically replaced with their row-based counterparts: the functions
\texttt{rowSums} and \texttt{rowMeans}. Thus, the expressions
``\texttt{sum(A{[}{[}1:Kmax{]}{]})}'' and
``\texttt{rowSums(A{[}{[}1:Kmax{]}{]})}'' can be used interchangeably.

We illustrate this syntax with a simple example. Suppose that we have a
\texttt{DAG} object, named ``\texttt{D}'' and we use the network
generator, named ``\texttt{rnet.gnp}''\footnote{Note that the network
  generator \texttt{rnet.gnp} is provided as part of the \pkg{simcausal}
  package. It uses the \pkg{igraph} \proglang{R} package function
  \texttt{sample\_gnp} to sample a network graph and then converts its
  output into the \pkg{simcausal} network matrix representation. See
  \texttt{?rnet.gnp} for additional information.}. As shown in the code
snippet below, we first define an empty \texttt{DAG} object \texttt{D},
then we add the network named ``\texttt{net}'' to object \texttt{D} and
we also define an IID Bernoulli variable named ``\texttt{Var}''.

\begin{Shaded}
\begin{Highlighting}[]
\KeywordTok{library}\NormalTok{(}\StringTok{"simcausal"}\NormalTok{)}
\NormalTok{D <-}\StringTok{ }\KeywordTok{DAG.empty}\NormalTok{() +}
\StringTok{  }\KeywordTok{network}\NormalTok{(}\StringTok{"net"}\NormalTok{, }\DataTypeTok{netfun =} \StringTok{"rnet.gnp"}\NormalTok{, }\DataTypeTok{p =} \FloatTok{0.1}\NormalTok{) +}
\StringTok{  }\KeywordTok{node}\NormalTok{(}\StringTok{"Var"}\NormalTok{, }\DataTypeTok{distr =} \StringTok{"rbern"}\NormalTok{, }\DataTypeTok{prob =} \FloatTok{0.5}\NormalTok{)}
\end{Highlighting}
\end{Shaded}

Next, we define a new node, named ``\texttt{Var.F1}'', as the value of
\texttt{Var} for the first friend of each observation \(i=1,\ldots,N\).
We do this by defining a node with a constant (degenerate) distribution,
\texttt{distr="rconst"}, and indexing the first friend of each
observation with the expression ``\texttt{Var{[}{[}1{]}{]}}'', as shown
in the following example:

\begin{Shaded}
\begin{Highlighting}[]
\NormalTok{D <-}\StringTok{ }\NormalTok{D +}\StringTok{ }\KeywordTok{node}\NormalTok{(}\StringTok{"Var.F1"}\NormalTok{, }\DataTypeTok{distr =} \StringTok{"rconst"}\NormalTok{, }\DataTypeTok{const =} \NormalTok{Var[[}\DecValTok{1}\NormalTok{]])}
\end{Highlighting}
\end{Shaded}

Suppose we now wish to list the \texttt{Var} values among the first 4
friends from each set \(F_i\). This can be accomplished by defining a
single multivariate node and using the expression
``\texttt{Var{[}{[}1:4{]}{]}}'', as shown next:

\begin{Shaded}
\begin{Highlighting}[]
\NormalTok{D <-}\StringTok{ }\NormalTok{D +}\StringTok{ }\KeywordTok{node}\NormalTok{(}\KeywordTok{paste0}\NormalTok{(}\StringTok{"Var.F"}\NormalTok{,}\DecValTok{1}\NormalTok{:}\DecValTok{4}\NormalTok{), }\DataTypeTok{distr =} \StringTok{"rconst"}\NormalTok{, }\DataTypeTok{const =} \NormalTok{Var[[}\DecValTok{1}\NormalTok{:}\DecValTok{4}\NormalTok{]])}
\end{Highlighting}
\end{Shaded}

As our final example, we define a new degenerate node, named
``\texttt{mean.F1to4}'', as the mean of \texttt{Var} amongst the first 4
friends, for each \(i=1,\ldots,N\). We also define a new Bernoulli node,
named ``\texttt{Var.F1to4}'', for which the \(i\)-specific probability
of success is also given as the mean of \texttt{Var} amongst the first 4
friends in \(F_i\). The data for the variables defined thus far can now
be simulated by simply calling the functions \texttt{set.DAG} and
\texttt{sim} in sequence and specifying the desired sample size \(N\)
with the argument ``\texttt{n}''.

\begin{Shaded}
\begin{Highlighting}[]
\NormalTok{D <-}\StringTok{ }\NormalTok{D +}
\StringTok{  }\KeywordTok{node}\NormalTok{(}\StringTok{"mean.F1to4"}\NormalTok{, }\DataTypeTok{distr =} \StringTok{"rconst"}\NormalTok{, }\DataTypeTok{const =} \KeywordTok{mean}\NormalTok{(Var[[}\DecValTok{1}\NormalTok{:}\DecValTok{4}\NormalTok{]], }\DataTypeTok{na.rm=}\OtherTok{TRUE}\NormalTok{)) +}
\StringTok{  }\KeywordTok{node}\NormalTok{(}\StringTok{"Var.F1to4"}\NormalTok{, }\DataTypeTok{distr =} \StringTok{"rbern"}\NormalTok{, }\DataTypeTok{prob =} \KeywordTok{mean}\NormalTok{(Var[[}\DecValTok{1}\NormalTok{:}\DecValTok{4}\NormalTok{]], }\DataTypeTok{na.rm=}\OtherTok{TRUE}\NormalTok{))}
\NormalTok{D <-}\StringTok{ }\KeywordTok{set.DAG}\NormalTok{(D, }\DataTypeTok{n.test =} \DecValTok{200}\NormalTok{)}
\KeywordTok{sim}\NormalTok{(D, }\DataTypeTok{n =} \DecValTok{50}\NormalTok{)}
\end{Highlighting}
\end{Shaded}

In summary, for a given indexing vector \texttt{indx}, the
network-indexing expression, such as ``\texttt{Var{[}{[}indx{]}{]}}'',
will evaluate to a matrix with \(N\) rows and the number of columns
determined by the length of \texttt{indx}. We note that indexing
variable \texttt{indx} can be a non-negative integer-valued vector, with
values starting from 0 and bounded above by a special reserved constant
named ``\texttt{Kmax}''. That is, the variable \texttt{Kmax} can be used
for finding out the maximal friend index for any given network, as we
demonstrate in examples in following sections. In addition, one can use
\texttt{0} as part of the same friend indexing vector, where the
expression ``\texttt{Var{[}{[}0{]}{]}}'' is equivalent to using
``\texttt{Var}''. This provides a convenient syntax for indexing the
actual \texttt{Var} value of observation \(i\) along with the
\texttt{Var} values of \(i\)'s friends, for example, allowing the
expressions such as ``\texttt{sum(Var{[}{[}0:Kmax{]}{]})}''.
Furthermore, for a specific observation \(i\), the expression
``\texttt{Var{[}{[}k{]}{]}}'' will evaluate to a missing value (i.e.,
``\texttt{NA}'') whenever \(i\) has fewer than \texttt{k} friends. This
default behavior can be also altered by passing a special named argument
``\texttt{replaceNAw0=TRUE}'' to the corresponding \texttt{node()} call,
in which case all of such missing (\texttt{NA}) values are automatically
replaced with \texttt{0} values. In addition, any node expression can
reference a special reserved variable ``\texttt{nF}'', which is a vector
of length \texttt{n} and it stores the total number of friends for
observation \(j\) in its \(j\)th entry.

Finally, we note that any function that defines a network-indexing node
summary can be similarly applied as a summary of a time-varying node and
vice-versa. For example, for a time varying node ``\texttt{Var.t}'', the
expression ``\texttt{sum(Var.t{[}t\_indx{]})}'' is analogous to our
previous example of the network summary
``\texttt{sum(Var{[}{[}indx{]}{]})}'', except that in the former case we
are summing the values of a time-varying node \texttt{Var.t} for
time-points defined by the indexing vector \texttt{t\_indx}. We also
provide some additional examples of such network summaries in the
following sections. For more in-depth description of this functionality
we refer to \citet{JSSsimcausal}, Section 2.3. ``\emph{Specifying a
structural equation model}''. Furthermore, both of these indexing
operators, i.e., the time indexing operator ``\texttt{{[}...{]}}'' and
the network indexing operator ``\texttt{{[}{[}...{]}{]}}'', can be
combined to form a single summary applied to a time-varying node, as we
demonstrate in Section \ref{sim.longdata}.

\section{Simulation Study Comparing Performance of Dependent-Data
Estimators}\label{simulation-study-comparing-performance-of-dependent-data-estimators}

\label{sim.Example1}

The simulation study described here is a simplified version of a
hypothetical observational study that might have been conducted to
evaluate the social influence of healthy living on personal long term
health status. We assumed that this study collected data on an
interconnected community of \(N\) individuals. For each individual
\(i\), we would have measured their social network \(F_i\), baseline
covariates \(W_i\), an exposure \(A_i\) and a binary outcome \(Y_i\).
The exposure was assessed as a continuous physical activity index and
the outcome indicated if the person was obese after some follow-up
period. We evaluated the average-causal effect of a stochastic
intervention that intervened only on some members of the community by
shifting their observed physical activity level by some known constant
\(shift>0\), while keeping the exposures of others unchanged. More
precisely, such stochastic intervention for each unit \(i\) can be
defined by a function \(\delta(A_i,W_i)\) that assigns the new
(intervened) exposure to either \(A_i+shift\) or \(A_i\), depending on
the covariate values \(W_i\). Such interventions have been described in
the past \citep{munoz2012population} and they arise naturally in
settings with continuous exposures where it is not feasible to intervene
on every member of the population. For example, in our hypothetical
study it might be infeasible to increase the level of physical activity
for an individual with a pre-existing medical condition who is above a
certain age. Therefore, as an alternative, we may consider a dynamic
interventions that does not intervene on such community members, as
determined by the pre-specified function \(\delta(A_i,W_i)\). This in
turn allows us to define the types of causal parameters which are less
likely to violate the positivity assumption \citep{robins1999} (also
known as the assumption of experimental treatment assignment (ETA)
\citep{neugebauer2005}).

In our simulation, we start by sampling a network graph for \(N\) units
(nodes) according to the ``small-world'' model
\citep{watts1998collective}, along with three unit-specific baseline
covariates, \[W_i = (W_i(1),W_i(2),W_i(3)),\] for \(i=1,\ldots,N\),
where \(W_i(1)\) is a categorical risk factor, and \((W_i(2),W_i(3))\)
are binary counfounders. The unit-specific exposure \(A_i\) is then
simulated conditionally on \(W_i\) as normal with mean
\(\mu(W_i)= 0.58*W_i(2)+0.33*W_i(3)\) and standard deviation
\(\sigma^2=1\). We also let the conditional density of \(A_i\) given
\(W_i\) to be denoted as \(g_0(a|w)\). We denote the intervened exposure
on \(i\) as \(A_i^*\), with its corresponding conditional density
denoted as \(g^*\). We set \(A^*_i\) to \(A_i+shift\), for some known
constant \(shift>0\), only if the following condition holds:

\begin{equation}
A_i - \mu(W_i) \leq \dfrac{\log(trunc)}{shift} + \dfrac{shift}{2}\label{eq:define-gstar}
\end{equation}

for known truncation constant \(trunc>0\), and otherwise, the
intervention keeps the observed exposure \(A_i\) unchanged. Note that
the above inequality in \ref{eq:define-gstar} is equivalent to the
following condition

\begin{equation}
\dfrac{g_0(A_i-shift|W_i)}{g_0(A_i|W_i)} \leq trunc,
\end{equation}

due to the fact that \(g_0\) is a normal density. Thus, by picking small
enough values of \(trunc\) and \(shift\), we can define interventions
that shift the observed values of \(A_i\) by \(shift\), for observations
that satisfy the inequality \ref{eq:define-gstar}. By defining our
intervention \(g^*\) in this manner we thus avoid interventions that
result in large values of \((g^{*}/g_0)(A_i|W_i)\), which reduces the
risk of violating the positivity assumption.

Finally, the binary outcome \(Y_i\) is simulated as dependent on all
three of \(i\)'s baseline covariates in \(W_i\), \(i\)'s exposure
\(A_i\), as well as the baseline covariate values and exposures of
\(i\)'s friends. In particular, we assume that the probability of
success for each binary outcome \(Y_i\) is logit-linear function of the
summary measures \((W_i^s,A_i^s)\), which are defined as
\[W_i^s:=(W_i, \dfrac{1}{|F_i|} \sum_{j \in F_i}{W_j(1)}, \sum_{j \in F_i}{W_j(2)}, \sum_{j \in F_i}{W_j(3)})\ \mbox{and}\ A_i^s:=(A_i, \sum_{j \in F_i}{A_j}).\]

Our target causal quantity is then defined as the sample-average of
\(N\) \(i\)-specific expected counterfactual outcomes \(Y^{*}_i\), under
stochastic intervention \(g^*\), for \(i=1,\ldots,N\). These
counterfactual outcomes are generated by modifying the observed
data-generating distribution, which consists of replacing the structural
equation for generating \(N\) observed exposures \(A_i\) with a new
structural equation that defines the intervened exposures \(A^*_i\), for
\(i=1,\ldots,N\). The outcomes \(Y^*_i\), for \(i=1,\ldots,N\) can then
be sampled from such a modified data-generating distribution. Note that
we are using the new notation \(Y^{*}_i\) to denote the fact that the
conditional distribution of the observed data \((W_i,A_i,Y_i)\) has been
modified by replacing \(g_0\) with the intervention \(g^*\). We denote
this causal quantity as \(\psi_0\) and define it as:
\[\psi_0 := E\left(\dfrac{1}{N} \sum_{i=1}^NY^*_i\right).\]

\subsection{Specifying the network distribution and the structural
equation model for dependent
data\label{sec:obsdat-sim1}}\label{specifying-the-network-distribution-and-the-structural-equation-model-for-dependent-data}

As noted, for this simulation the network graph is sampled according to
the small-world network model \citep{watts1998collective}. We will use
the network generator function ``\texttt{rnet.SmWorld}'', which is
provided by the \pkg{simcausal} package. This function serves as a
wrapper for the function \texttt{sample\_smallworld} of the \pkg{igraph}
\proglang{R} package \citep{igraph}, which performs the actual sampling
of an undirected network graph. The nodes of the graph returned by the
\texttt{sample\_smallworld} function are treated as individual
observations and are indexed as \(i=1,\ldots,N\). An undirected edge
between two nodes \(i\) and \(j\) is interpreted as two directed edges,
meaning that unit \(j\) is a friend of unit \(i\) and vice versa. In
terms of our notation, this means that \(j \in F_i\) and \(i \in F_j\).
The resulting network object produced by the \pkg{igraph} package is
then converted into its \pkg{simcausal} network representation, where
each friend set \(F_i\) is determined according to the above described
rule. We remind the reader that \pkg{simcausal} represents the network
as a matrix with \(N\) rows, each row \(i\) representing the set of
friends \(F_i\) (social connections of unit \(i\)), where each \(F_i\)
may be padded with extra \texttt{NA}s to make sure all \(F_i\) have the
same length.

We proceed by first instantiating an empty \texttt{DAG} object, which is
assigned to variable ``\texttt{D}'' below. We will use this \texttt{DAG}
object to define the network distribution, as well as to encode the
unit-specific distribution of the data. As a next step, we register the
network generator function (\texttt{rnet.SmWorld}), making it a part of
the object \texttt{D}. This will allow us to: (a) generate a specific
network, and (b) use this network as a way to connect nodes defined
within the same \texttt{DAG} object \texttt{D}. Specifically, as we show
in the following code snippet, we add a network to the object \texttt{D}
by using syntax ``\texttt{D\textless{}-D+network()}''. The name of the
network generator is passed on to the \texttt{network()} call with an
argument ``\texttt{netfun}''. The \texttt{network()} call also requires
a ``\texttt{name}'' argument (first argument), which specifies a name of
a particular network\footnote{The actual value of the \texttt{name}
  argument of \texttt{network()} function becomes only relevant when the
  user wants to either over-write the existing network or define several
  different networks within the same \texttt{DAG} object.}. The function
\texttt{network} allows passing any number of additional arguments,
which will be passed to the underlying network generator function (e.g.,
arguments \texttt{dim}, \texttt{nei} and \texttt{p} in the example
below). These additional arguments allow the user to further parametrize
the distribution of the network graph.

\begin{Shaded}
\begin{Highlighting}[]
\NormalTok{D <-}\StringTok{ }\KeywordTok{DAG.empty}\NormalTok{()}
\NormalTok{D <-}\StringTok{ }\NormalTok{D +}\StringTok{ }\KeywordTok{network}\NormalTok{(}\StringTok{"Net"}\NormalTok{, }\DataTypeTok{netfun =} \StringTok{"rnet.SmWorld"}\NormalTok{, }\DataTypeTok{dim =} \DecValTok{1}\NormalTok{, }\DataTypeTok{nei =} \DecValTok{3}\NormalTok{, }\DataTypeTok{p =} \FloatTok{0.3}\NormalTok{)}
\end{Highlighting}
\end{Shaded}

Next, we define a 3-dimensional IID covariate \(W=(W(1),W(2),W(3))\),
where \(W(1)\) is a categorical node ``\texttt{W1}'', \(W(2)\) is a
binary node ``\texttt{W2}'' and \(W(3)\) is a binary node
``\texttt{W3}'', as shown below:

\begin{Shaded}
\begin{Highlighting}[]
\NormalTok{D <-}\StringTok{ }\NormalTok{D +}
\StringTok{  }\KeywordTok{node}\NormalTok{(}\StringTok{"W1"}\NormalTok{, }\DataTypeTok{distr =} \StringTok{"rcat.b0"}\NormalTok{, }\DataTypeTok{probs =} \KeywordTok{c}\NormalTok{(}\FloatTok{0.0494}\NormalTok{, }\FloatTok{0.1823}\NormalTok{, }\FloatTok{0.2806}\NormalTok{, }\FloatTok{0.2680}\NormalTok{, }\FloatTok{0.1651}\NormalTok{, }\FloatTok{0.0546}\NormalTok{)) +}
\StringTok{  }\KeywordTok{node}\NormalTok{(}\StringTok{"W2"}\NormalTok{, }\DataTypeTok{distr =} \StringTok{"rbern"}\NormalTok{, }\DataTypeTok{prob =} \KeywordTok{plogis}\NormalTok{(-}\FloatTok{0.2} \NormalTok{+}\StringTok{ }\NormalTok{W1/}\DecValTok{3}\NormalTok{)) +}
\StringTok{  }\KeywordTok{node}\NormalTok{(}\StringTok{"W3"}\NormalTok{, }\DataTypeTok{distr =} \StringTok{"rbern"}\NormalTok{, }\DataTypeTok{prob =} \FloatTok{0.6}\NormalTok{)}
\end{Highlighting}
\end{Shaded}

In our next step, we define the conditional distribution of a continuous
exposure \(A_i\) (node ``\texttt{A.obs}''), which is sampled as normal,
with its mean given by a linear combination of \(W_i(2)\) and \(W_i(3)\)
(nodes ``\texttt{W2}'' and ``\texttt{W3}'') and the standard deviation
of 1. The following example also defines another node named
``\texttt{A}'', which is set equal to \texttt{A.obs}. The node
\texttt{A} plays no role in the definition of our observed data
generating process. However, it will be utilized when we define our
intervention \(g^{*}\) in Section \ref{sec:intervene-sim1}. By
intervening on \texttt{A}, instead of \texttt{A.obs}, we will be able to
define an intervention~that is itself a function of the observed
exposure values stored in \texttt{A.obs}.

\begin{Shaded}
\begin{Highlighting}[]
\NormalTok{D <-}\StringTok{ }\NormalTok{D +}
\StringTok{  }\KeywordTok{node}\NormalTok{(}\StringTok{"A.obs"}\NormalTok{, }\DataTypeTok{distr =} \StringTok{"rnorm"}\NormalTok{, }\DataTypeTok{mean =} \FloatTok{0.58} \NormalTok{*}\StringTok{ }\NormalTok{W2 +}\StringTok{ }\FloatTok{0.33} \NormalTok{*}\StringTok{ }\NormalTok{W3, }\DataTypeTok{sd =} \DecValTok{1}\NormalTok{)+}
\StringTok{  }\KeywordTok{node}\NormalTok{(}\StringTok{"A"}\NormalTok{, }\DataTypeTok{distr =} \StringTok{"rconst"}\NormalTok{, }\DataTypeTok{const =} \NormalTok{A.obs)}
\end{Highlighting}
\end{Shaded}

Finally, for each unit \(i=1,\ldots,N\), we model the conditional
probability of success for the binary outcome \(Y_i\) (node
``\texttt{Y}'') as a logit-linear function of the earlier described
summary measures \((W_{i}^s,A_{i}^s)\). These summaries include
\((W_i,A_i)\), as well as the covariate summaries of units in \(F_i\)
(i.e., variables measured on friends of unit \(i\)). In more detail, for
each unit \(i\), these network summaries include the average of
\(W_j(1)\) and the sum of the exposures \(A_j\) among all units \(j\)
that are friends with \(i\) (i.e., \(j \in F_i\)).

\begin{Shaded}
\begin{Highlighting}[]
\NormalTok{D <-}\StringTok{ }\NormalTok{D +}
\StringTok{  }\KeywordTok{node}\NormalTok{(}\StringTok{"Y"}\NormalTok{, }\DataTypeTok{distr =} \StringTok{"rbern"}\NormalTok{,}
    \DataTypeTok{prob =} \KeywordTok{plogis}\NormalTok{(}\DecValTok{5} \NormalTok{+}\StringTok{ }\NormalTok{-}\FloatTok{0.5}\NormalTok{*W1 -}\StringTok{ }\FloatTok{0.58}\NormalTok{*W2 -}\StringTok{ }\FloatTok{0.33}\NormalTok{*W3 +}
\StringTok{                  }\NormalTok{-}\FloatTok{1.5}\NormalTok{*}\KeywordTok{ifelse}\NormalTok{(nF >}\StringTok{ }\DecValTok{0}\NormalTok{, }\KeywordTok{sum}\NormalTok{(W1[[}\DecValTok{1}\NormalTok{:Kmax]])/nF, }\DecValTok{0}\NormalTok{) +}
\StringTok{                  }\NormalTok{-}\FloatTok{1.4}\NormalTok{*}\KeywordTok{sum}\NormalTok{(W2[[}\DecValTok{1}\NormalTok{:Kmax]]) +}\StringTok{ }\FloatTok{2.1}\NormalTok{*}\KeywordTok{sum}\NormalTok{(W3[[}\DecValTok{1}\NormalTok{:Kmax]]) +}
\StringTok{                  }\NormalTok{+}\FloatTok{0.35}\NormalTok{*A +}\StringTok{ }\FloatTok{0.15}\NormalTok{*}\KeywordTok{sum}\NormalTok{(A[[}\DecValTok{1}\NormalTok{:Kmax]])}
                  \NormalTok{),}
    \DataTypeTok{replaceNAw0 =} \OtherTok{TRUE}\NormalTok{)}
\NormalTok{Dset <-}\StringTok{ }\KeywordTok{set.DAG}\NormalTok{(D, }\DataTypeTok{n.test =} \DecValTok{200}\NormalTok{)}
\end{Highlighting}
\end{Shaded}

Note that the above object \texttt{Dset} has saved the described
data-generating distribution, which includes the definition of the
network and that of unit-specific data. In other words, \texttt{Dset}
object saves all the information that is needed in order to be able to
simulate: (1) the network graph on \(N\) units, and (2) the observations
\((W_i,A_i,Y_i)\), for \(i=1,\ldots,N\). Also note that such
unit-specific data may or may not be dependent (i.e., it may or may not
use the network structure), depending on a particular SEM
parameterization selected by the user.

\subsection{Simulating network and observed
data}\label{simulating-network-and-observed-data}

In our next code example we simulate the observed data on \(N\) units
from the distribution specified in the above object \texttt{Dset}. This
is accomplished by calling the function \texttt{sim}, where the total
number of units \(N\) is specified with the argument \texttt{n}. This
example saves the resulting data frame with observations
\((W_i,A_i,Y_i)\), for \(i=1,\ldots,N\), as an object named
``\texttt{datO}''. The network matrix and the vector counting the number
of friends for each unit are saved as objects named
``\texttt{NetInd\_mat}'' and ``\texttt{nF}'', respectively.

\begin{Shaded}
\begin{Highlighting}[]
\NormalTok{nsamp <-}\StringTok{ }\DecValTok{100}
\NormalTok{datO <-}\StringTok{ }\KeywordTok{sim}\NormalTok{(Dset, }\DataTypeTok{n =} \NormalTok{nsamp, }\DataTypeTok{rndseed =} \DecValTok{54321}\NormalTok{)}
\NormalTok{NetInd_mat <-}\StringTok{ }\KeywordTok{attributes}\NormalTok{(datO)$netind_cl$NetInd}
\NormalTok{nF <-}\StringTok{ }\KeywordTok{attributes}\NormalTok{(datO)$netind_cl$nF}
\end{Highlighting}
\end{Shaded}

The \texttt{plot.igraph} function in \pkg{igraph} package can be used
for visualizing such simulated network. However, the network ID matrix
\texttt{NetInd\_mat} needs to be first converted back into its original
\pkg{igraph} object representation (\texttt{g}), as we show below.

\begin{Shaded}
\begin{Highlighting}[]
\KeywordTok{library}\NormalTok{(}\StringTok{"igraph"}\NormalTok{)}
\NormalTok{g <-}\StringTok{ }\KeywordTok{sparseAdjMat.to.igraph}\NormalTok{(}\KeywordTok{NetInd.to.sparseAdjMat}\NormalTok{(NetInd_mat, }\DataTypeTok{nF =} \NormalTok{nF))}
\end{Highlighting}
\end{Shaded}

\begin{figure}[H]

{\centering \includegraphics{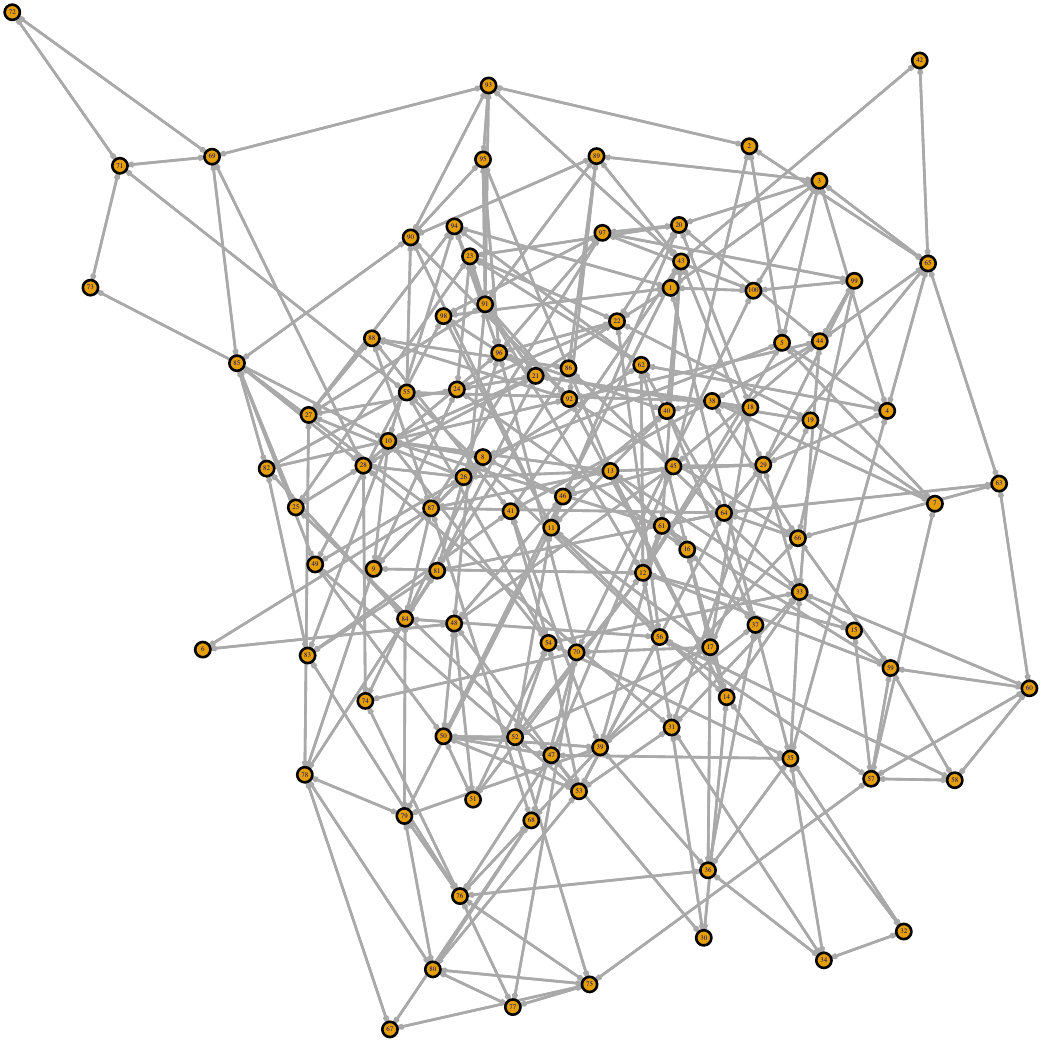}

}

\caption{Example of a network sampled from the small-world network model.\label{fig:netExample1}}\label{fig:netplot}
\end{figure}

We can now apply the \texttt{plot.igraph} function of the \pkg{igraph}
package to visualize the simulated network structure stored in the
object \texttt{g}, as shown in Figure \ref{fig:netExample1}.

\subsection{Defining interventions, simulating counterfactual data and
evaluating the true value of the causal
quantity\label{sec:intervene-sim1}}\label{defining-interventions-simulating-counterfactual-data-and-evaluating-the-true-value-of-the-causal-quantity}

Our next goal is to evaluate the true value of the causal parameter
\(\psi_0\) under intervention \(g^*\) via Monte-Carlo simulation that
samples from the distribution of the counterfactual outcomes \(Y_i^*\),
for \(i=1,\ldots,N\). This process consists of two stages. First, as we
show in the following code snippet, the intervention of interest is
defined by modifying some of the nodes of the observed data-generating
distribution. This is accomplished by calling the function
\texttt{action} and \emph{adding} the result of this call to the
previously defined \texttt{DAG} object \texttt{Dset}. The resulting
\emph{modified} structural equation model defines the distribution of
the counterfactual data, which includes the counterfactual outcome.
Importantly, each action-specific counterfactual distribution is saved
within the same object, along with the observed data generating
distribution, allowing the user to define any number of such
interventions. Below we define an intervention named ``\texttt{gstar}''
by providing a new definition of the exposure node \texttt{A}. This
intervention is defined according to our previously described stochastic
intervention \(g^*\), using the assignment criteria in equation
\ref{eq:define-gstar}. Note that the definition of \(g^*\) requires the
input of the observed unit-specific exposures, and in our current
example these exposure values were defined in node \texttt{A.obs}.

\begin{Shaded}
\begin{Highlighting}[]
\NormalTok{Dset <-}\StringTok{ }\NormalTok{Dset +}
\StringTok{  }\KeywordTok{action}\NormalTok{(}\StringTok{"gstar"}\NormalTok{,}
    \DataTypeTok{nodes =} \KeywordTok{node}\NormalTok{(}\StringTok{"A"}\NormalTok{, }\DataTypeTok{distr =} \StringTok{"rconst"}\NormalTok{,}
      \DataTypeTok{const =} \KeywordTok{ifelse}\NormalTok{(A.obs -}\StringTok{ }\NormalTok{(}\FloatTok{0.58}\NormalTok{*W2 +}\StringTok{ }\FloatTok{0.33}\NormalTok{*W3) >}\StringTok{ }\NormalTok{(}\KeywordTok{log}\NormalTok{(trunc)/shift +}\StringTok{ }\NormalTok{shift/}\DecValTok{2}\NormalTok{),}
                     \NormalTok{A.obs,}
                     \NormalTok{A.obs +}\StringTok{ }\NormalTok{shift)),}
    \DataTypeTok{trunc =} \DecValTok{1}\NormalTok{, }\DataTypeTok{shift =} \FloatTok{0.3}\NormalTok{)}
\end{Highlighting}
\end{Shaded}

Second, we perform Monte-Carlo simulation by sampling from this newly
defined counterfactual distribution defined by action \texttt{gstar} and
generate the counterfactual data, as shown in the following code
example. We do this by calling the function \texttt{sim} and specifying
the name of the previously defined action with an argument
\texttt{actions}. That is, the procedure in \texttt{sim} first samples
the network of size \(N=500\), proceeded by sampling of the
counterfactual data. Note that our target causal quantity \(\psi_0\) is
defined conditionally on the network sample size \(N\), namely, as an
average of \(N\) expected counterfactual outcomes. This implies that the
true value of \(\psi_0\) is allowed to change as the network's sample
size changes. Therefore, in the following evaluation of \(\psi_0\) we
use fixed sample size (\(N=500\)), which allows us to obtain Monte-Carlo
estimates of true causal quantity \(\psi_0:=E[1/N \sum_{i=1}^NY^*_i]\).
In particular, a single Monte-Carlo simulation step consists of sampling
\(N\) counterfactual outcomes \(Y_i^*\) and then taking the empirical
mean of these outcomes over 500 observations, thus yielding a single
estimate of \(\psi_0\). By taking the mean over 10,000 of such
Monte-Carlo estimates we can obtain a very close approximation of
\(\psi_0\). This causal parameter now defines our \emph{gold standard}:
the quantity which we may use for evaluating and comparing the
performance of different statistical methods. One possible application
of such causal parameters is exhibited in the following section.

\begin{Shaded}
\begin{Highlighting}[]
\KeywordTok{require}\NormalTok{(}\StringTok{"doParallel"}\NormalTok{)}
\KeywordTok{registerDoParallel}\NormalTok{(}\DataTypeTok{cores =} \KeywordTok{detectCores}\NormalTok{())}
\NormalTok{networksize <-}\StringTok{ }\DecValTok{500}
\NormalTok{nsims_psi0 <-}\StringTok{ }\DecValTok{10000}
\NormalTok{psi0_reps <-}\StringTok{ }\KeywordTok{foreach}\NormalTok{(}\DataTypeTok{i.sim =} \KeywordTok{seq}\NormalTok{(nsims_psi0), }\DataTypeTok{.combine =} \StringTok{"c"}\NormalTok{) %dopar%}\StringTok{ }\NormalTok{\{}
  \NormalTok{dat_gstar <-}\StringTok{ }\KeywordTok{sim}\NormalTok{(Dset, }\DataTypeTok{actions=}\StringTok{"gstar"}\NormalTok{, }\DataTypeTok{n =} \NormalTok{networksize)[[}\StringTok{"gstar"}\NormalTok{]]}
  \NormalTok{psi0 <-}\StringTok{ }\KeywordTok{mean}\NormalTok{(dat_gstar[[}\StringTok{"Y"}\NormalTok{]])}
\NormalTok{\}}
\NormalTok{psi0 <-}\StringTok{ }\KeywordTok{mean}\NormalTok{(psi0_reps)}
\KeywordTok{print}\NormalTok{(psi0)}
\end{Highlighting}
\end{Shaded}

\begin{verbatim}
## [1] 0.743815
\end{verbatim}

\subsection{Case study: comparing performance of dependent-data
estimators}\label{case-study-comparing-performance-of-dependent-data-estimators}

\label{sim.case.study}

We describe one possible application for the \pkg{simcausal} package by
conducting a simulation study that compares the performance of the three
dependent-data estimators implemented in \pkg{tmlenet} \proglang{R}
package \citep{R-tmlenet}. The data is simulated according to the above
described data generating distribution. The estimation procedures
implemented in \pkg{tmlenet} have been previously described in
\citet{sofryginJCI2016}. In particular, \pkg{tmlenet} package implements
three dependent-data estimators: Target Maximum Likelihood Estimator
(TMLE), the Inverse Probability Weighted Estimator (IPW) and the
G-computation substitution estimator (GCOMP). Our intervention of
interest and the corresponding target causal quantity were introduced in
the previous section. We will now use this target causal quantity as our
\emph{gold standard}, namely, it is the quantity that will generally
remain known in a real data-generating process. Our simulation study
hence uses this gold standard as a mean of comparing the finite sample
performance of these three dependent-data estimators over iterated
samples of hypothetical observed data. For additional details on these
estimators we refer to \citep{sofryginJCI2016}.

As our first step, we simulate the observed data, as shown below, for
500 units.

\begin{Shaded}
\begin{Highlighting}[]
\NormalTok{datO <-}\StringTok{ }\KeywordTok{sim}\NormalTok{(Dset, }\DataTypeTok{n =} \DecValTok{500}\NormalTok{)}
\NormalTok{net_obj <-}\StringTok{ }\KeywordTok{attributes}\NormalTok{(datO)[[}\StringTok{"netind_cl"}\NormalTok{]]}
\NormalTok{NetInd_mat <-}\StringTok{ }\NormalTok{net_obj[[}\StringTok{"NetInd"}\NormalTok{]]}
\NormalTok{nF <-}\StringTok{ }\NormalTok{net_obj[[}\StringTok{"nF"}\NormalTok{]]}
\NormalTok{Kmax <-}\StringTok{ }\NormalTok{net_obj[[}\StringTok{"Kmax"}\NormalTok{]]}
\end{Highlighting}
\end{Shaded}

As our next step, we define the input parameters for the
\texttt{tmlenet()} estimation function, as shown in the following code
snippets. In particular, we define the observed-data baseline summary
measures \(W_i^s\) by calling the \pkg{tmlenet} function
``\texttt{def\_sW}'', as shown below. We also define the exposure
summary measures \(A_i^s\) by calling the \pkg{tmlenet} function
``\texttt{def\_sA}'', as shown below. The summary measures defined by
\texttt{def\_sW} and \texttt{def\_sA} will be utilized by \pkg{tmlenet}
for constructing estimators of \(\psi_0\). Note that the functions
\texttt{def\_sW} and \texttt{def\_sA} use syntax that is nearly
identical to the previously described \pkg{simcausal} syntax for
specifying the same network-based summaries.

\begin{Shaded}
\begin{Highlighting}[]
\KeywordTok{require}\NormalTok{(}\StringTok{"tmlenet"}\NormalTok{)}
\NormalTok{def.sW <-}\StringTok{ }\KeywordTok{def_sW}\NormalTok{(W1, W2, W3) +}
\StringTok{          }\KeywordTok{def_sW}\NormalTok{(}\DataTypeTok{meanW1 =} \KeywordTok{ifelse}\NormalTok{(nF >}\StringTok{ }\DecValTok{0}\NormalTok{, }\KeywordTok{sum}\NormalTok{(W1[[}\DecValTok{1}\NormalTok{:Kmax]])/nF, }\DecValTok{0}\NormalTok{), }\DataTypeTok{replaceNAw0 =} \OtherTok{TRUE}\NormalTok{) +}
\StringTok{          }\KeywordTok{def_sW}\NormalTok{(}\DataTypeTok{sumW2 =} \KeywordTok{sum}\NormalTok{(W2[[}\DecValTok{1}\NormalTok{:Kmax]]), }\DataTypeTok{replaceNAw0 =} \OtherTok{TRUE}\NormalTok{) +}
\StringTok{          }\KeywordTok{def_sW}\NormalTok{(}\DataTypeTok{sumW3 =} \KeywordTok{sum}\NormalTok{(W3[[}\DecValTok{1}\NormalTok{:Kmax]]), }\DataTypeTok{replaceNAw0 =} \OtherTok{TRUE}\NormalTok{)}
\NormalTok{def.sA <-}\StringTok{ }\KeywordTok{def_sA}\NormalTok{(A, }\DataTypeTok{sumA =} \KeywordTok{sum}\NormalTok{(A[[}\DecValTok{1}\NormalTok{:Kmax]]), }\DataTypeTok{replaceNAw0 =} \OtherTok{TRUE}\NormalTok{)}
\end{Highlighting}
\end{Shaded}

In the following example we define the intervention of interest by
calling the \pkg{tmlenet} function ``\texttt{def\_new\_sA}''. We again
note the syntactic similarities between the intervention specification
in the example below and the specification of the counterfactual action
\texttt{gstar} in the example from the previous section.

\begin{Shaded}
\begin{Highlighting}[]
\NormalTok{trunc <-}\StringTok{ }\DecValTok{1}\NormalTok{; shift <-}\StringTok{ }\FloatTok{0.5}
\NormalTok{newA.gstar <-}\StringTok{  }\KeywordTok{def_new_sA}\NormalTok{(}\DataTypeTok{A =} \KeywordTok{ifelse}\NormalTok{(A -}\StringTok{ }\NormalTok{(}\FloatTok{0.58}\NormalTok{*W2 +}\StringTok{ }\FloatTok{0.33}\NormalTok{*W3) >}\StringTok{ }\NormalTok{(}\KeywordTok{log}\NormalTok{(trunc)/shift +}\StringTok{ }\NormalTok{shift/}\DecValTok{2}\NormalTok{),}
                          \NormalTok{A,}
                          \NormalTok{A +}\StringTok{ }\NormalTok{shift))}
\end{Highlighting}
\end{Shaded}

In our following example we define the regression formulas, where the
regression defined by the variable ``\texttt{Qform}'' is used for
modeling the conditional outcome \(Y_i\) given the previously defined
summary measures \((W_i^s,A_i^s)\) and the regression in
``\texttt{hform}'' is used for modeling the conditional probabilities of
the observed exposure summaries \(A_i^s\) given the observed baseline
summaries \(W_i^s\), namely, \texttt{hform} describes the model for
\(P(A_i^s|W_i^s)\). We also call \texttt{tmlenet\_options} to specify
additional tuning parameters for the \pkg{tmlenet}. In particular, we
specify the parameter ``\texttt{bin.method}'', which defines the method
for discretizing continuous variables in \(A_i^s\) (those variables are
detected automatically by \pkg{tmlenet}). Furthermore, we specify the
parameter ``\texttt{maxNperBin}'', which defines the maximum number of
bins (cutoff intervals) when performing such discretization (see
\texttt{?tmlenet\_options} for additional details and for other optional
tuning parameters available in \pkg{tmlenet}).

\begin{Shaded}
\begin{Highlighting}[]
\NormalTok{Qform <-}\StringTok{ "Y ~ A + sumA + meanW1 + sumW2 + sumW3 + W1 + W2 + W3"}
\NormalTok{hform <-}\StringTok{ "A + sumA ~ meanW1 + sumW2 + sumW3 + W1 + W2 + W3"}
\KeywordTok{tmlenet_options}\NormalTok{(}\DataTypeTok{bin.method =} \StringTok{"equal.mass"}\NormalTok{, }\DataTypeTok{maxNperBin =} \DecValTok{50}\NormalTok{)}
\end{Highlighting}
\end{Shaded}

Finally, we call the estimation routine \texttt{tmlenet()}, as shown
below, where we pass all of the above-defined variables, including the
previously simulated dependent data \texttt{datO}, as well as the
network matrix \texttt{NetInd\_mat}. Furthermore, we specify some
optional parameters by using the argument \texttt{optPars}. In
particular, by passing a list with \texttt{bootstrap.var = TRUE}, we
request that the TMLE inference is performed with parametric bootstrap
(along with two standard approaches described below). Finally, the
argument \texttt{n.bootstrap} controls the number of the parametric
bootstrap samples (higher number will lead to more precise bootstrap
variance estimates, but will significantly increase the computation
time).

\begin{Shaded}
\begin{Highlighting}[]
\NormalTok{res <-}\StringTok{ }\KeywordTok{tmlenet}\NormalTok{(}\DataTypeTok{data =} \NormalTok{datO, }\DataTypeTok{sW =} \NormalTok{def.sW, }\DataTypeTok{sA =} \NormalTok{def.sA,}
                \DataTypeTok{Kmax =} \NormalTok{Kmax, }\DataTypeTok{NETIDmat =} \NormalTok{NetInd_mat,}
                \DataTypeTok{intervene1.sA =} \NormalTok{newA.gstar ,}
                \DataTypeTok{Qform =} \NormalTok{Qform, }\DataTypeTok{hform.g0 =} \NormalTok{hform,}
                \DataTypeTok{optPars =} \KeywordTok{list}\NormalTok{(}\DataTypeTok{bootstrap.var =} \OtherTok{TRUE}\NormalTok{, }\DataTypeTok{n.bootstrap =} \DecValTok{100}\NormalTok{))}
\end{Highlighting}
\end{Shaded}

The above function call returns a list containing the estimates of the
parameter \(\psi_0\), where \(\psi_0\) is defined as the sample-average
of the expected counterfactual outcomes under the intervention specified
in \texttt{newA.gstar}. The table of estimates consists of the TMLE, IPW
and GCOMP and it can be obtained by running the following code (output
not shown):

\begin{Shaded}
\begin{Highlighting}[]
\NormalTok{res[[}\StringTok{"EY_gstar1"}\NormalTok{]][[}\StringTok{"estimates"}\NormalTok{]]}
\end{Highlighting}
\end{Shaded}

The \texttt{tmlenet} function also returns the corresponding 95\%
confidence intervals (CIs) for the TMLE and IPW under the list item
named \texttt{"IC.CIs"}. Note that these CIs are based on the variance
estimates of the influence curves that also adjusts for the
network-based dependence among the units, as described in
\citet{sofryginJCI2016}. These CIs can be printed by running the
following code (output not shown):

\begin{Shaded}
\begin{Highlighting}[]
\NormalTok{res[[}\StringTok{"EY_gstar1"}\NormalTok{]][[}\StringTok{"IC.CIs"}\NormalTok{]]}
\end{Highlighting}
\end{Shaded}

Furthermore, an alternative estimate of the 95\% CI for the TMLE can be
obtained via the parametric bootstrap approach. This approach also
estimates the TMLE variance in the presence of network-based dependence
among units. However, in small samples the parametric bootstrap can
achieve better performance (better nominal coverage), compared to the
above described influence curve-based variance estimate of the TMLE. The
corresponding CI for the parametric bootstrap is stored under the list
item named \texttt{"boot.CIs"} and it can be obtained by running the
following code (output not shown):

\begin{Shaded}
\begin{Highlighting}[]
\NormalTok{res[[}\StringTok{"EY_gstar1"}\NormalTok{]][[}\StringTok{"boot.CIs"}\NormalTok{]]}
\end{Highlighting}
\end{Shaded}

Finally, one can also obtain the 95\% CIs for the TMLE and IPW based on
the approach that estimates the variance of corresponding influence
curves assuming the observed units are IID. One would expect that such
IID-based CIs are going to be overly optimistic (too narrow) when, in
fact, the units are dependent. Thus, this approach will generally result
in CIs with inadequate coverage, as is also demonstrated by our
simulation study below. Finally, these IID-based 95\% CIs can be
obtained by running the following code (output not shown):

\begin{Shaded}
\begin{Highlighting}[]
\NormalTok{res[[}\StringTok{"EY_gstar1"}\NormalTok{]][[}\StringTok{"iid.CIs"}\NormalTok{]]}
\end{Highlighting}
\end{Shaded}

Our simulation study repeated the above described estimation procedure
5,000 times, each time using a newly sampled dataset of \(500\)
network-dependent observations. The corresponding \proglang{R} code for
this simulation study can be found in Supplementary Materials. We
evaluated the absolute bias, the mean-squared error (MSE) and the
variance of the above described three estimators, with results presented
in Table \ref{simrestab} (all performance metrics were multiplied by
10). In addition, we evaluated the mean of the TMLE variance estimate
based on the parametric bootstrap procedure, along with its
corresponding 95\% CI, as reported in columns ``BOOT.VAR'' and
``BOOT.CI.cover'' of Table \ref{simrestab.cover}. We also evaluated the
mean of the TMLE variance estimate which adjusted for the dependence
among units based on the estimated variance of the efficient influence
curve and the coverage of its corresponding 95\% CI, as reported in
columns ``DEP.VAR'' and ``DEP.CI.cover'' in Table \ref{simrestab.cover}.
Finally, we evaluated the mean of the TMLE variance estimate which
completely ignored the network information and assumed units are IID,
along with its corresponding 95\% CI, as reported in columns ``IID.VAR''
and ``IID.CI.cover'' of Table \ref{simrestab.cover}. As expected, the
IID-based variance estimator resulted in 95\% CIs that had inadequate
coverage, compared to the coverage probabilities of 0.935 and 0.924 for
the parametric bootstrap and the influence curve-based CIs,
respectively.

\begin{table}[H]
\begin{center}
\begin{tabular}{lrrr}
\toprule
\multicolumn{1}{c}{Estimator}&\multicolumn{1}{c}{Bias10}&\multicolumn{1}{c}{MSE10}&\multicolumn{1}{c}{Variance10}\tabularnewline
\midrule
TMLE&0.0042&0.01371&0.01372\tabularnewline
IPW&0.0183&0.01724&0.01721\tabularnewline
GCOMP&0.0019&0.01278&0.01278\tabularnewline
\bottomrule
\end{tabular}

\caption{Performance of the three dependent data estimators across $5,000$ simulations, each simulation consisted of $N=500$ network-dependent units. The reported bias, mean squared error (MSE) and variance are multiplied by 10.\label{simrestab}}\end{center}

\end{table}

\begin{table}[H]
\begin{center}
\begin{tabular}{rrrrrr}
\toprule
\multicolumn{1}{c}{BOOT.VAR}&\multicolumn{1}{c}{DEP.VAR}&\multicolumn{1}{c}{IID.VAR}&\multicolumn{1}{c}{BOOT.CI.cover}&\multicolumn{1}{c}{DEP.CI.cover}&\multicolumn{1}{c}{IID.CI.cover}\tabularnewline
\midrule
0.001354&0.001265&0.000506&0.935&0.924&0.756\tabularnewline
\bottomrule
\end{tabular}

\caption{Monte-Carlo approximated mean of the TMLE variance based on parametric bootstrap ('BOOT.VAR'), dependent data ('DEP.VAR'), TMLE variance estimator for IID data ('IID.VAR'), the coverage of the 95\% CI for parametric bootstrap variance ('BOOT.CI.cover'), dependent data variance ('DEP.CI.cover') and IID variance ('IID.CI.cover') estimates across $5,000$ simulations, each simulation consisting of $N=500$ network-dependent units.\label{simrestab.cover}}\end{center}

\end{table}

\subsection{Case study: the effect of network dependence on TMLE
inference}\label{case-study-the-effect-of-network-dependence-on-tmle-inference}

\label{sim.case.study.spillover}

We also conducted a simulation study where we varied the strength of the
effect of the above~defined baseline network summaries
\(((1/|F_i|) \sum_{j \in F_i}{W_j(1)}, \sum_{j \in F_i}{W_j(2)}, \sum_{j \in F_i}{W_j(3)})\)
on the outcome \(Y_i\) of each unit \(i\). In particular, we considered
9 data-generating scenarios, where in each scenario these three network
summaries had different coefficient values in the formula that defines
the distribution of the node \texttt{Y}. The first simulation (Scenario
1) was based on a data-generating distribution with coefficient values
of (-0.5, -0.4, -0.1), thus, resulting in the following definition of
node \texttt{Y}:

\begin{Shaded}
\begin{Highlighting}[]
\KeywordTok{node}\NormalTok{(}\StringTok{"Y"}\NormalTok{, }\DataTypeTok{distr =} \StringTok{"rbern"}\NormalTok{,}
  \DataTypeTok{prob =} \KeywordTok{plogis}\NormalTok{(}\DecValTok{5} \NormalTok{+}\StringTok{ }\NormalTok{-}\FloatTok{0.5}\NormalTok{*W1 -}\StringTok{ }\FloatTok{0.58}\NormalTok{*W2 -}\StringTok{ }\FloatTok{0.33}\NormalTok{*W3 +}
\StringTok{                }\NormalTok{-}\FloatTok{0.5}\NormalTok{*}\KeywordTok{ifelse}\NormalTok{(nF >}\StringTok{ }\DecValTok{0}\NormalTok{, }\KeywordTok{sum}\NormalTok{(W1[[}\DecValTok{1}\NormalTok{:Kmax]])/nF, }\DecValTok{0}\NormalTok{) +}
\StringTok{                }\NormalTok{-}\FloatTok{0.4}\NormalTok{*}\KeywordTok{sum}\NormalTok{(W2[[}\DecValTok{1}\NormalTok{:Kmax]]) +}\StringTok{ }\NormalTok{-}\FloatTok{0.1}\NormalTok{*}\KeywordTok{sum}\NormalTok{(W3[[}\DecValTok{1}\NormalTok{:Kmax]]) +}
\StringTok{                }\NormalTok{+}\FloatTok{0.35}\NormalTok{*A +}\StringTok{ }\FloatTok{0.15}\NormalTok{*}\KeywordTok{sum}\NormalTok{(A[[}\DecValTok{1}\NormalTok{:Kmax]])}
                \NormalTok{),}
  \DataTypeTok{replaceNAw0 =} \OtherTok{TRUE}\NormalTok{)}
\end{Highlighting}
\end{Shaded}

The last simulation (Scenario 9) used the coefficient values of (-1.5,
-1.4, 2.1), i.e., the values used in the original definition of node
\texttt{Y}, as defined above in Section \ref{sec:obsdat-sim1}. Finally,
the simulation scenarios in between (Scenario 2 - Scenario 8) set these
coefficient values at equally spaced intervals between those used in
Scenario 1 and Scenario 9. Thus, the outcome \(Y_i\) in Scenario 1 could
be described as being the least dependent on the baseline covariate
values of \(i\)'s friends (baseline network summaries), while the
simulations in Scenario 2 to 9 progressively increase the strength of
this dependence. We also kept the intervention and the parameter of
interest unchanged from the previous simulation. Furthermore, we
assessed the effects of varying the strength of this relationship on
TMLE inference, as shown in Table \ref{simrestab.spillover}. As
expected, the coverage probability of the IID-based 95\% CIs that ignore
between-unit dependence (column ``IID.CI.cover'') decreases as we
continue to increase the dependence of each \(Y_i\) on the baseline
covariates of \(i\)'s friends.

\begin{table}[H]
\begin{center}
\begin{tabular}{llrrrrr}
\toprule
\multicolumn{1}{l}{sims}&\multicolumn{1}{c}{BOOT.VAR}&\multicolumn{1}{c}{DEP.VAR}&\multicolumn{1}{c}{IID.VAR}&\multicolumn{1}{c}{BOOT.CI.cover}&\multicolumn{1}{c}{DEP.CI.cover}&\multicolumn{1}{c}{IID.CI.cover}\tabularnewline
\midrule
Scenario 1&0.000954&0.000858&0.000689&0.934&0.901&0.878\tabularnewline
Scenario 2&0.000706&0.000642&0.000503&0.930&0.894&0.875\tabularnewline
Scenario 3&0.000773&0.000703&0.000499&0.936&0.909&0.866\tabularnewline
Scenario 4&0.000600&0.000548&0.000370&0.938&0.911&0.859\tabularnewline
Scenario 5&0.000682&0.000626&0.000387&0.935&0.913&0.844\tabularnewline
Scenario 6&0.000767&0.000712&0.000402&0.939&0.916&0.827\tabularnewline
Scenario 7&0.000750&0.000694&0.000367&0.937&0.913&0.814\tabularnewline
Scenario 8&0.001231&0.001152&0.000498&0.935&0.924&0.769\tabularnewline
Scenario 9&0.001353&0.001275&0.000508&0.938&0.925&0.747\tabularnewline
\bottomrule
\end{tabular}

\caption{The effect of network dependence on inference across $5,000$ simulations, each simulation consisting of $N=500$ dependent units. The results are ordered by increasing strength of network dependence of $Y_i$ on the baseline covariates of $i$'s friends. The first row corresponds with a simulation that has the least dependence, with subsequent rows ordered by the increasing amount of the dependence.\label{simrestab.spillover}}\end{center}

\end{table}

\section{Discussion}\label{discussion}

We described how the \pkg{simcausal} \proglang{R} package can facilitate
the conduct of network-based simulation studies in causal inference
research, specifically allowing one to model data with known network and
known functional form of dependence among units. We also described how
the package can be used for simulations with longitudinal data
structures. We described how the \pkg{simcausal} \proglang{R} package
allows creating a wide range of artificial datasets often encountered in
public health applications of causal inference methods. In particular,
this includes simulations of causal mechanisms of interference or
spillover. To simplify the specification of such models we implemented a
novel \proglang{R} syntax which allows for a concise and intuitive
expression of complex functional dependencies for a large number of
nodes. We also demonstrated how this syntax can be used for simplifying
the specification and simulation of complex network-based summaries of
the data. Moreover, we argued that such complex simulations are often
necessary when one tries to conduct a realistic simulation study that
attempts to replicate a large variety of scenarios one might expect to
see from a true data-generating process. We also note that our package
can work in conjunction with other network simulation tools, as we
demonstrated with the example of the \pkg{igraph} \proglang{R} package
that was used to sample the actual networks. In addition, the
\pkg{simcausal} package allows the user to specify and simulate
counterfactual data under various interventions (e.g., static, dynamic,
deterministic, or stochastic). These interventions may represent
exposures to treatment regimens, the occurrence or non-occurrence of
right-censoring events, or of specific monitoring events. These
simulations provide practical settings with the types of data generating
distributions which might be used for validation, testing and comparison
of statistical methods for causal inference. To the best of our
knowledge there are no other tools which implement the specific type of
functionality afforded by the network-based syntax that is implemented
in \pkg{simcausal} and is described in this article.

One of the distinguishing features of \pkg{simcausal} is that it allows
the user to define and compute various causal target parameters, such
as, the treatment-specific counterfactual mean, that can then serve as
the model-free \emph{gold standard}. That is, the causal parameter is
always the same functional of the counterfactual data distribution,
regardless of the user-selected parameterization of the structural
equation model. For example, the gold standard defined in this manner
provides an objective measure of bias that does not depend on the
modeling assumptions of a specific statistical method. Coupled with a
wide variety of possible data generating distributions that may be
specified in \pkg{simcausal}, this package provides statisticians with a
powerful tool for testing the validity and accuracy of various
statistical methods. For example, one may use our package for validating
an implementation of a novel statistical method, using the simulated
data with the known truth (the true value of the causal parameter),
prior to applying such an algorithm to real data, in which this truth is
unknown. As another example, one may use \pkg{simcausal} to simulate
data from a variety of data-generating distributions and conduct a
simulation study comparing the properties of different statistical
procedures (e.g., bias, mean-squared error (MSE), asymptotic confidence
interval coverage) against the user-selected causal parameter. We
emphasize that the main purpose of our package is not to assess the
impact of real-life interventions, but rather to test the validity and
performance of statistical methods, which can then be applied to real
datasets. Moreover, we demonstrated that the \pkg{simcausal}
\proglang{R} package is a flexible tool that enables easier
communication of assumptions between various practitioners and thus
helps improve the transparency about the assumptions of different
statistical methods.

We also note that the implementation of additional features in future
releases of the \pkg{simcausal} package should further expand its
utility for methods research. Among such possible improvements is to
allow interventions on the network structure and provide a unified
interface for changes to friend structure as part of the intervention.
Future work will also focus on modeling the changes in network structure
over time, for example, by providing an interface for specifying a
time-varying analogs of the \code{network} function, modeling the
probability of forming a new tie or removing a certain friend over time
and allowing one to sample new networks conditional on the previously
sampled networks.

\section{Acknowledgments}\label{acknowledgments}

This research was supported by NIH grant R01 AI074345-07.

\bibliography{SimCausal_Networks_2016.bib}

\end{document}